%% file: main.tex
\documentclass[lettersize,journal]{IEEEtran}
\input{setup}

\IEEEoverridecommandlockouts

\begin{document}


\title{Enhancing LLM-based Fault Localization with a Functionality-Aware Retrieval-Augmented Generation Framework}

\author{Xinyu Shi\orcidlink{0009-0005-1836-6822}, 
        Zhenhao Li\orcidlink{0000-0002-4909-1535}, 
        and An Ran Chen\orcidlink{0000-0003-3137-7540}
\thanks{Xinyu Shi and An Ran Chen are with the Department of Electrical and Computer Engineering, University of Alberta, Edmonton, Canada (e-mail: xshi12@ualberta.ca; anran6@ualberta.ca).}%
\thanks{Zhenhao Li is with the School of Information Technology, York University, Toronto, Canada (e-mail: lzhenhao@yorku.ca).}
}


\maketitle

\input{sections/abstract}


\input{sections/introduction}

\input{sections/relatedwork}

\input{sections/methodology}

\input{sections/experiment}

\input{sections/results}

\input{sections/discussion}

\input{sections/threats}

\input{sections/conclusion}

\bibliographystyle{IEEEtran}

\bibliography{references}

\end{document}

%% file: setup.tex
\usepackage{algorithmic}
\usepackage{graphicx}
\usepackage{textcomp}
\usepackage{xcolor}
\usepackage{xspace}
\usepackage{multirow}
\usepackage{booktabs}
\usepackage{makecell}

\usepackage{xspace}

\usepackage[most]{tcolorbox}
\tcbset{
  leftlineonly/.style={
    enhanced,
    breakable,
    colback=gray!8,  
    boxrule=0pt,             
    colframe=white,           
    arc=8pt,                  
    left=6pt, right=4pt, top=4pt, bottom=4pt
  }
}

\newcommand{\fixme}[1]{\textcolor{orange}{{\it [fixme]}}}

\newcommand{\rqboxc}[2]{%
  \begin{tcolorbox}[
    enhanced,
    colback=white,
    colframe=black,
    coltitle=white,
    fonttitle=\bfseries,
    title=#1,
    attach boxed title to top left,
    boxed title style={
      colframe=black,
      colback=black,
      size=small,       
      sharp corners=south
    },
    sharp corners,
    boxrule=1pt,
    left=4pt, right=4pt, top=4pt, bottom=4pt
  ]
  #2
  \end{tcolorbox}%
}

\newcommand{\tool}{FaR-Loc\xspace}

\newcommand{\phead}[1]{\noindent {\bf #1}}
\newcommand{\uhead}[1]{\noindent {\underline {#1}}}

\usepackage{listings}

\usepackage{orcidlink}

%% file: sections/abstract.tex
\begin{abstract}
Fault localization (FL) is a critical but time-consuming task in software debugging, aiming to identify faulty code elements. While recent advances in large language models (LLMs) have shown promise for FL, they often struggle with complex systems due to the lack of project-specific knowledge and the difficulty of navigating large projects. To address these limitations, we propose FaR-Loc, a novel framework that enhances method-level FL by integrating LLMs with retrieval-augmented generation (RAG). FaR-Loc consists of three key components: LLM Functionality Extraction, Semantic Dense Retrieval, and LLM Re-ranking. First, given a failed test and its associated stack trace, the LLM Functionality Extraction module generates a concise natural language description that captures the failing behavior. Next, the Semantic Dense Retrieval component leverages a pre-trained code-understanding encoder to embed both the functionality description (natural language) and the covered methods (code) into a shared semantic space, enabling the retrieval of methods with similar functional behavior. Finally, the LLM Re-ranking module reorders the retrieved methods based on their contextual relevance. Our experiments on the widely used Defects4J benchmark show that FaR-Loc outperforms state-of-the-art LLM-based baselines SoapFL and AutoFL, by 14.6\% and 9.1\% in Top-1 accuracy, by 19.2\% and 22.1\% in Top-5 accuracy, respectively. It also surpasses all learning-based and spectrum-based baselines across all Top-N metrics without requiring re-training. 
Furthermore, we find that pre-trained code embedding models that incorporate code structure, such as UniXcoder, can significantly improve fault localization performance by up to 49.0\% in Top-1 accuracy.
Finally, we conduct a case study to illustrate the effectiveness of \tool and to provide insights for its practical application.
\end{abstract}

%% file: sections/introduction.tex
\section{Introduction}


\noindent Fault localization (FL) is a fundamental but time-consuming task in software debugging, with the goal of identifying faulty code elements. Research indicates that more than half of the debugging time is spent on finding the fault location \cite{bohmeWhereBugHow2017, alaboudi2021exploratorystudydebuggingepisodes,chen2021demystifying}. To reduce manual effort, researchers have introduced various automated FL techniques. Among these, spectrum-based~\cite{abreu2006evaluation,abreu2007accuracy,wong2013dstar} and learning-based techniques~\cite{wong2014boosting, sohn2017fluccs, li2019deepfl} have gained popularity due to their effectiveness.

Recent advancements in Large Language Models (LLMs) have introduced new opportunities for improving FL. In particular, these models have been pre-trained on large corpus of code and textual data, which provides them with strong capability in code comprehension and reasoning~\cite{nam2024using,chen2025reasoning}.
However, applying LLMs to fault localization presents several key challenges.
First, LLMs often operate with limited debugging information, such as stack traces and test failures. Without additional insights, the model may fail to capture the full scope of the fault and tend to hallucinate~\cite{rafi2024enhancing,xuFlexFLFlexibleEffective2024}. Second, LLMs must navigate the entire codebase, where identifying relevant context for the fault becomes difficult. Efficient retrieval techniques are essential to prioritize the most fault-relevant information, but current approaches often struggle on large-scale codebases~\cite{wuLargeLanguageModels2023,yang2025empirical}. Finally, reasoning over complex code structures remains a significant challenge. LLMs may lack the deep semantic understanding necessary to track how faults propagate in complex systems, which can degrade their localization accuracy~\cite{qinFaultLocalizationSemantic2024,qinSoapFLStandardOperating2025,yangLargeLanguageModels2024}.

Recent LLM-based fault localization methods explore various strategies to support code navigation within large projects. 
Some methods narrow down the fault location by letting LLMs make step-wise decisions. One common strategy is hierarchical navigation~\cite{jiang2025cosil, yu2025orcaloca, chen2025locagent, qinSoapFLStandardOperating2025, qinFaultLocalizationSemantic2024}, where the LLM first identifies suspicious files or classes before narrowing down to specific methods. Another approach provides a set of external functions for LLMs to call~\cite{kangQuantitativeQualitativeEvaluation2024}, such as retrieving a specific class or accessing comments. However, they are prone to cascading errors: an early misstep in the retrieval process (e.g., selecting the wrong file or class) can propagate into subsequent errors, particularly when LLMs are required to make decisions under limited contextual information.
Other approaches use LLM-generated code summaries at varying granularities~\cite{yeo2025improving, qinSoapFLStandardOperating2025, qinFaultLocalizationSemantic2024}.
Although these summaries can condense context and potentially facilitate localization, they may introduce substantial computational overhead, increase the likelihood of hallucinations and error accumulation, and omit critical information like root cause of the fault.
These trade-offs underscore the need for more efficient FL techniques that support robust code navigation while reducing dependence on large-scale summaries and high-stakes LLM decision-making.

Recently, retrieval-augmented approaches have been proposed for various downstream software engineering tasks, such as code generation~\cite{li2023skcoder,zhou2022docprompting}, test case generation~\cite{shin2024retrieval,arora2024generating}, log analysis~\cite{ma2024librelog,zhang2024lograg}, and automated program repair~\cite{wang2023rap,nashid2023retrieval}. These approaches enhance the capabilities of LLMs by incorporating task-relevant context through retrieval, all without requiring model re-training. 
Despite their demonstrated success, retrieval-augmented techniques have not yet been explored in the context of fault localization. 
In line with this paradigm, our intuition is that \textit{the test failure information (e.g., stack traces) can be augmented as failing functionality, which then serves as a retrieval query for identifying suspicious methods and reasoning about them}. This retrieval-augmented formulation narrows the search space to a semantically and functionally relevant subset of the codebase, thereby reducing input complexity.


In this paper, we propose a novel framework, \tool (\textbf{F}unctionality-\textbf{A}ware fault \textbf{loc}alization Framework via \textbf{R}etrieval-Augmented LLMs), which leverages a retrieval augmented framework to improve performance in large-context scenarios. 
\tool consists of three main components: \textbf{LLM Functionality Extraction}, \textbf{Semantic Dense Retrieval}, and \textbf{LLM Re-ranking}.
Specifically, the LLM Functionality Extraction module first augments the sparse debugging information (i.e., stack traces) into a functionality query. This helps to drastically reduce the irrelevant or noisy debugging information.
Then, the functionality query is passed to the Semantic Dense Retrieval component, which performs the retrieval process to extract only the most relevant methods to the functionality query. This step helps reduce the large search space to prioritize the most suspicious covered methods.
Finally, the LLM Re-ranking module re-ranks the identified suspicious methods by reasoning around the failing functionality. This approach enables the LLM to reason about the fault using a concise and focused query, rather than being overwhelmed by excessive test failure information.



We evaluate the effectiveness of \tool
on the widely adopted Defects4J-V1.2.0 dataset~\cite{just2014defects4j}, which includes 395 real-world faults across eight studied systems. The results show that \tool consistently outperforms both the state-of-the-art LLM-based and learning-based baselines.
Specifically, \tool outperforms state-of-the-art LLM-based baselines SoapFL~\cite{qinSoapFLStandardOperating2025} and AutoFL~\cite{kangQuantitativeQualitativeEvaluation2024}, by 14.6\% and 9.1\% in Top-1 accuracy, by 19.2\% and 22.1\% in Top-5 accuracy, respectively. It also surpasses all learning-based and spectrum-based baselines across all Top-N metrics.
Our cost analysis also shows that \tool is competitively efficient.
We further conduct an ablation study to assess the contribution of individual component within \tool. 
Our results show that when integrating with the re-ranking mechanism, \tool can filter out the noisy candidates through semantic reasoning, achieving a 24.8\% improvements in Top-5 accuracy.
We also analyze the impact of different LLMs and embedding models on \tool's performance to understand how the choice of model affects each component.
In addition, we conduct a case study to illustrate the rationale behind \tool's effectiveness.
Finally, to assess the generalizability of our approach, we further evaluate \tool on 226 additional faults from the Defects4J-V2.0.0 dataset~\cite{defects4j}. Across this extended evaluation, \tool continues to outperform all baselines, which demonstrates its robustness.
The main contributions of this paper are:

\begin{itemize}
  \item We propose a novel LLM-based fault localization framework, \tool, that combines Large Language Model (LLM) with Retrieval-Augmented Generation (RAG). \tool retrieves suspicious methods by embedding both the failing functionality and covered methods into a shared semantic space. It then locates the most suspicious methods by re-ranking the retrieved methods based on their contextual relevance to the failing functionality.
  \item Our extensive evaluation shows that \tool outperforms state-of-the-art LLM-based baselines SoapFL~\cite{qinSoapFLStandardOperating2025} and AutoFL~\cite{kangQuantitativeQualitativeEvaluation2024}, by 14.6\% and 9.1\% in Top-1 accuracy, by 19.2\% and 22.1\% in Top-5 accuracy, respectively. It also surpasses all learning-based~\cite{wong2014boosting, sohn2017fluccs} and spectrum-based~\cite{abreu2007accuracy} baselines across all Top-N metrics.
  \item We observe that while the functionality queries generated by different LLMs are largely consistent, their ability to re-rank suspicious methods varies significantly.
  \item We observe that code embeddings vary in their effectiveness for FL. Embedding models that incorporate Abstract Syntax Trees (ASTs) during pre-training, such as UniXcoder~\cite{guoUniXcoderUnifiedCrossModal2022}, can boost Top-1 accuracy by up to 49\%.
  \item We conduct a case study to evaluate the effectiveness of \tool on complex systems. Our finding shows that functionality queries provide additional semantic context, semantic dense retrieval identifies relevant methods, and LLM re-ranking further refines the candidate list.
\end{itemize}

%% file: sections/relatedwork.tex
\section{Background \& Related Work}

\subsection{Fault Localization}

\phead{Spectrum-Based and Learning-Based Fault Localization.}
Prior to the emergence of LLMs, a variety of automated fault localization techniques have been developed. Spectrum-Based Fault Localization (SBFL) methods compute suspiciousness scores based on the coverage spectra information of passing and failing test cases \cite{abreu2007accuracy},  but their accuracy remains limited in practice \cite{pearson2017evaluating,wenHistoricalSpectrumBased2019,kochhar2016practitioners}.
Learning-based Fault Localization (LBFL) utilize machine learning algorithms to learn the fault patterns from previous bugs \cite{wong2014boosting,niu2025deep} or integrates multiple metrics to rank potential fault locations \cite{b2016learning, sohn2017fluccs, li2019deepfl,chakraborty2024rlocator}. Although effective in many scenarios, LBFL often require large amounts of labeled data and extensive training. They typically depend on manually crafted features, risk overfitting to specific projects, and struggle to generalize to previously unseen systems. Moreover, they lack the ability to provide explanations for the potential causes of faults.

\phead{LLM-based Fault Localization.}
Recent advancements in Large Language Models (LLMs) have introduced new opportunities for improving fault localization.
Early LLM-based fault localization research primarily concentrated on constrained contexts, such as identifying buggy lines within individual files or methods. 
Wu et al.~\cite{wuLargeLanguageModels2023} prompt ChatGPT-3.5/4 with buggy methods and error logs for line-level fault localization, but observe that performance deteriorate significantly even when the context is merely expanded to the class level.
LLMAO~\cite{yangLargeLanguageModels2024} fine-tunes bidirectional adapters on LLMs to identify buggy lines, but is limited to fixed-length code segments within 128 lines and cannot scale to project-level localization.

Recent LLM-based techniques employ different strategies to navigate entire codebases, enabling fault localization at the project level.
Some approaches incorporate traditional fault localization techniques or predefined helper functions to support code navigation.
FlexFL~\cite{xuFlexFLFlexibleEffective2024} relies on traditional FL techniques to narrow the search space before invoking LLMs, making its performance heavily dependent on the quality of initial retrieval.
AutoFL~\cite{kangQuantitativeQualitativeEvaluation2024} equips LLMs with function-calling to navigate the codebase via externally defined functions. However, it is constrained by a fixed function call budget, and at each step, the model must make decisions based on limited information such as class and method signatures.
Other approaches adopt a coarse-to-fine localization strategy, first narrowing down to files or modules and then to specific functions, often leveraging LLMs to generate textual summaries of code for auxiliary guidance. 
CosFL~\cite{qinFaultLocalizationSemantic2024} clusters code into modules based on the call graph and uses LLMs to generate summaries for each module and method, enabling hierarchical localization from modules to methods. 
SoapFL~\cite{qinSoapFLStandardOperating2025} employs a multi-agent architecture that performs file-level localization followed by method-level localization, and further utilizes LLMs to complete missing documentation comments. 
However, when dealing with complex systems, these approaches may suffer from error propagation across stages—if the initial step (such as class or module localization) is incorrect, subsequent efforts will be misguided. Moreover, these methods rely heavily on LLM-generated textual summaries or documentation, which may exhibit hallucinations or omit error-relevant details.


Our proposed framework, \tool, effectively leverages the functionality information of failing behaviors and employs retrieval-augmented generation (RAG) techniques to significantly reduce the candidate list presented to the LLM. This design not only enables the LLM to deeply reason about the failure context, but also mitigates the risk of hallucination accumulation and error propagation introduced by repeated and redundant LLM invocations.

\subsection{Retrieval-Augmented Generation}
\noindent Retrieval-Augmented Generation (RAG) is a paradigm that enhances LLMs by integrating external retrieval mechanisms~\cite{gao2024retrievalaugmented,shuster2021retrieval,chen2024benchmarking}. Unlike traditional retrieval approaches, RAG dynamically incorporates information from external sources such as documentation, codebases, or knowledge bases. This architecture provides three advantages: First, it substantially reduces hallucination by grounding model outputs in factual, retrievable information, thereby improving response reliability. Second, it enables models to leverage information beyond their pre-training cutoff, accessing specialized knowledge without requiring re-training. Finally, RAG frameworks facilitate continuous knowledge updates through simple modifications to the retrieval corpus, allowing systems to remain up-to-date with evolving information without model redeployment. These capabilities make RAG particularly valuable for domains with rapidly changing information or specialized knowledge requirements. 
Although RAG has been successfully applied to a variety of downstream software engineering tasks, including code suggestion~\cite{chen2024code,yang2025empirical,xu2025mantra}, code optimization\cite{gao2024search}, and automated program repair~\cite{wang2023rap,nashid2023retrieval},
to the best of our knowledge, its integration with LLM-based fault localization has not yet been explored.

\subsection{Code Embeddings}

\noindent Code embeddings are dense vector representations of code snippets that encode their semantic and structural information~\cite{ding2022can}. Early techniques, such as bag-of-words and TF-IDF, treat code as plain text and fail to capture its hierarchical and syntactic nature. Later neural approaches like code2vec~\cite{code2vec} and code2seq~\cite{code2seq} improve this by leveraging syntax paths to model code structure, yet they remain limited in handling long-range dependencies and generalizing across languages.

Transformer-based models have greatly expanded the application of code embeddings by enabling richer, context-aware representations of code. 
Models such as CodeBERT~\cite{fengCodeBERTPreTrainedModel2020} and GraphCodeBERT~\cite{guo2020graphcodebert} jointly learn from source code and natural language using masked language modeling and structural objectives. Later architectures, including UniXcoder, further advanced this paradigm by incorporating multiple code modalities, while CodeT5~\cite{wang2021codet5} and CodeT5+~\cite{wangCodeT5OpenCode2023} unified understanding and generation tasks within a single framework.

A key innovation in modern embedding models is their ability to align code and natural language within a shared representation space, facilitating cross-modal tasks such as retrieval and reasoning. In this work, we focus on the representation learning capability of these models, specifically their encoder-side embeddings, which are sufficient for downstream applications like semantic code search and fault localization.

%% file: sections/methodology.tex
\section{Methodology}

\begin{figure*}[t]
  \centering
  \includegraphics[width=\textwidth]{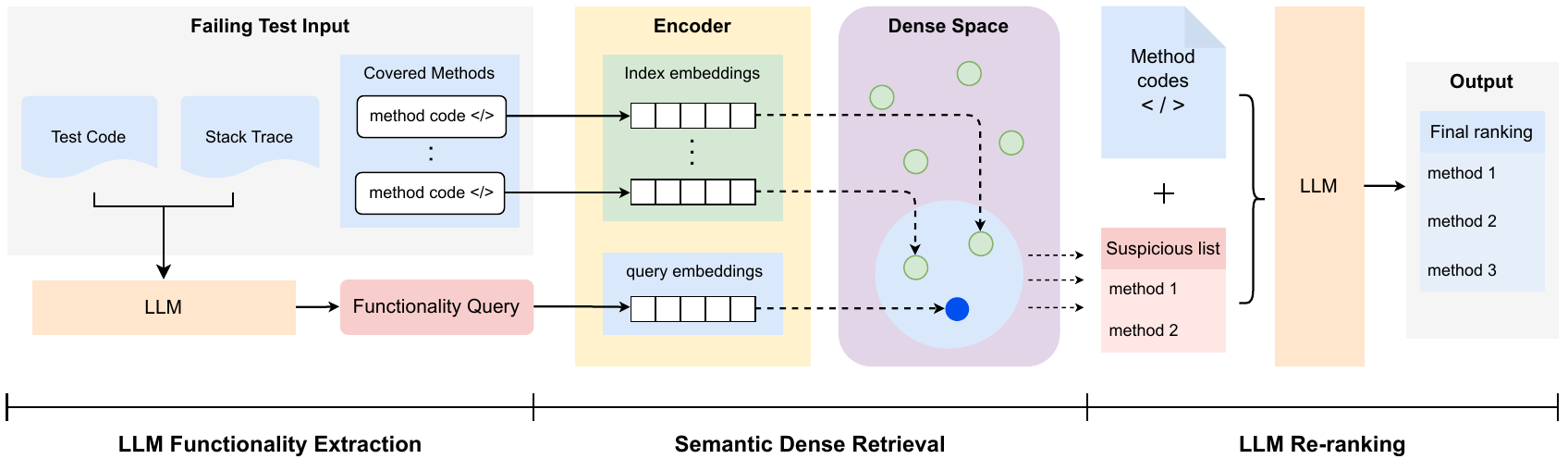}
  \caption{Overview of the \tool framework.}
  \label{fig:overview}
\end{figure*}

\noindent We propose \tool, a LLM-based framework for method-level fault localization.
\tool leverages a retrieval-augmented framework that retrieves code snippets semantically similar to the failing functionality to help in fault understanding. The core intuition behind \tool is that by first understanding which functionality is failing, we can use that insight to more precisely locate the corresponding faults in the source code.
At a high level, \tool takes failed tests and its associated stack trace as inputs to generate the failing functionality, which then serves as a search query to find suspicious methods.
Figure~\ref{fig:overview} shows an overview of our framework, which comprises three components: (1) \textbf{LLM Functionality Extraction}, (2) \textbf{Semantic Dense Retrieval}, and (3) \textbf{LLM Re-ranking}.
First, \tool identifies the failing functionality through test code and stack trace. It leverages the LLM's reasoning capabilities to generate a natural language description of what functionality is broken.
Then, \tool narrows down potential faulty methods by using the semantic dense retrieval mechanism to identify code segments that align with the failing functionality. The objective of this retrieval process is to pinpoint the methods relevant to the root cause of the fault.
Lastly, we refine the overall ranking of the suspicious methods list by leveraging an LLM-based re-ranking mechanism. 
This step further improves the accuracy of the ranked list by re-evaluating the relevance of each method in the context of the failing functionality.

\subsection{LLM Functionality Extraction}

\noindent In this component, we utilize the LLM to analyze the failing test information and generate a concise description of the underlying functionality. 
Prior studies have shown that LLMs are well-suited for tasks such as code context understanding~\cite{nam2024using} and reasoning runtime behaviours~\cite{chen2025reasoning}. Therefore, we leverage the LLM's advanced reasoning capabilities to generate a high-level functionality description to center the debugging process around the intent of failing code.
The output of this component is a natural language description, which is then employed as an augmented query for the semantic dense retrieval stage.

The LLM Functionality Extraction component addresses the challenge of limited debugging information. Although test failure information (i.e., stack traces, test results) has been shown to be useful in LLM-based FL~\cite{wuLargeLanguageModels2023,kangLargeLanguageModels2023}, LLMs can still struggle to identify the fault due to the limited nature of this information. For instance, stack traces typically contain only a narrow snapshot of the program's state at the time of failure, primarily showing the call stack rather than the full execution context. As such, they offer a structural view of the failure but lack contextual and semantic details. To mitigate this limitation, we generate a functionality query that provides additional semantic context, clarifying the intended behavior of the code. This not only improves the LLM's understanding of the failure but also filters out noise from irrelevant information, such as third-party, built-in, or test helper method calls commonly found in stack traces. As a result, LLM can better focus on the core failing behavior when locating faults.

To effectively guide the LLM in uncovering the underlying functionality information behind failing tests, we carefully design the prompt tailored for the fault localization task.
Motivated by recent research in LLM prompt engineering~\cite{cheng2024prompt} that shows role-playing can effectively stimulate the LLM's knowledge in specific domains, we assign LLM the role of a code assistant.
This prompt provides the LLM with the failing test code snippets and stack traces, and encourages it to reason about the functionality that is failing.
The prompt is shown below.

\begin{tcolorbox}[
  colback=gray!10, 
  colframe=black!50,
  title=Prompt for LLM Functionality Extraction,
  left=1mm,        
  right=1mm,       
  top=1mm,         
  bottom=1mm       
]
\small
You are a code assistant helping to identify faulty program behavior. One or more unit tests have failed due to the same underlying functionality issue.
\\
Given the following test failure information (including multiple test codes, and stack traces), extract **only** the underlying functional logic that failed. Your output should be a clean, concise description of the shared functionality that failed to be implemented correctly.
\\

Requirements:

- Focus on what functionality failed, not how the tests failed.

- Include any relevant objects, inputs, and expected behavior if available.

- The description should be precise and suitable for use as a semantic query to retrieve code (in natural language).

- Avoid unrelated details. \\

Test name: \{test name\}\\
Test code: \{test code\}\\
Stack trace: \{stack trace\}
\end{tcolorbox}

To ensure the LLM is provided with sufficient context, we incorporate detailed failing test information, including both the test code snippets and corresponding stack traces. Specifically, we extract the fully qualified names of the failing tests from the test framework and statically locate the associated code snippets within the source files. In cases where multiple failing tests share identical test code due to inheritance from a common test class, we include only the parent test method in the prompt to avoid redundancy. Additionally, the stack traces from the failing tests are included to offer further contextual insights into the failure, enabling the LLM to generate a more precise and comprehensive functionality description.

\subsection{Semantic Dense Retrieval}

\noindent In this component, we leverage the failing functionality description and the failing test coverage as input. Specifically, we use the functionality description (natural language) as the query and the methods (code) covered by the failing tests as the index to identify the most relevant methods. We first determine the methods covered by the failing tests using Cobertura~\cite{cobertura}, a dynamic Java coverage tool integrated into the Defects4J dataset framework. Next, we extract the code snippets of these covered methods and construct an index for the retrieval process. The objective is to retrieve methods that are semantically aligned with the failing functionality description, thereby improving the likelihood of finding the actual source of the bug.

To bridge the gap between natural language descriptions and code representations, we leverage pre-trained code-understanding encoders—such as CodeBERT, CodeT5+, and UniXcoder—that embed both natural language and code into a shared semantic space. These models are specifically trained to capture cross-modal semantic relationships~\cite{fengCodeBERTPreTrainedModel2020, wangCodeT5OpenCode2023,guoUniXcoderUnifiedCrossModal2022}, allowing the functionality description and method implementations to be meaningfully compared during retrieval. 
For the small subset of method snippets that exceed the token limit of the code embedding models, we divide each method into smaller chunks that fit within the limit and compute embeddings for each chunk. To obtain a single representation for the method, we aggregate these embeddings using max pooling. This approach is motivated by the observation that different segments of a method may contain important semantic information (e.g., parameter lists, return statements), and max pooling effectively preserves the most salient features across all chunks.
After encoding, the functionality description is treated as a query, while the code snippets of covered methods form the retrieval index. We utilize FAISS~\cite{douze2024faiss} to construct an efficient index that supports scalable nearest-neighbor search. Cosine similarity~\cite{salton1975vector} is used to compute semantic relevance between the query and indexed methods, and the top-k most relevant candidates are retrieved.

Our Semantic Dense Retrieval module narrows the input scope for the LLM by retrieving methods that are semantically aligned with the failing behavior. Using a cross-modal encoder, it effectively captures functional similarities between code and natural language, addressing code and structural complexity. Prior approaches\cite{qinSoapFLStandardOperating2025,zhang2024autocoderover,kangQuantitativeQualitativeEvaluation2024} often depend on limited contextual cues (e.g., file or class names) or generate synthetic documentation via another LLM, which can introduce hallucinations or overlook the root cause of the failure. In contrast, by focusing on functionality, our approach reduces context size while minimizing hallucination risk.

\subsection{LLM Re-ranking}

\noindent In this component, the LLM re-ranks the suspicious methods identified in the previous stage. It leverages the failing functionality description produced by the LLM Functionality Extraction module, together with the code snippets of the retrieved candidate methods, to generate a final ranked list of suspicious methods. By providing both the high-level functionality description and the detailed method implementations, the LLM can more accurately understand the relationship between the failing behavior and specific code, which is often essential for method-level localization~\cite{qinSoapFLStandardOperating2025}.

Prior approaches~\cite{kangQuantitativeQualitativeEvaluation2024, qinSoapFLStandardOperating2025} often require the LLM to make decisions (e.g., invoke external function calls or perform coarse-grained localization) while being overwhelmed by excessive failure information and lacking project-specific knowledge. They typically involve multiple rounds of reasoning, increasing the risk of hallucinations and error propagation. In contrast, our LLM Re-ranking module enables decision-making over a filtered set of candidate methods that are functionally relevant to the failure, significantly reducing input complexity. This focused scope allows full code snippets to be included within context limits. Moreover, our functionality query helps the LLM better understand the failing behavior, which further improves the localization accuracy.

The output of this module is a ranked list of faulty methods, where the most likely buggy methods are positioned at the top. To ensure clarity and consistency, the LLM is explicitly instructed to rank the methods in descending order of their likelihood of being faulty, with rank 1 assigned to the most suspicious method. 

We also prompt LLM to generate the results in a structured JSON format, which includes the class name, method name, and rank for each method. This format not only simplifies integration with downstream fault localization tasks but also enhances the interpretability and usability of the results. 
The prompt used for this re-ranking step is carefully designed to guide the LLM in performing an accurate and thorough evaluation. It explicitly instructs the LLM to analyze each method's code snippet in the context of the failing functionality and to produce a ranked list in the specified JSON format.
The prompt is shown below.

\begin{tcolorbox}[
  colback=gray!10, 
  colframe=black!50,
  title=Prompt for LLM Re-ranking,
  left=1mm,        
  right=1mm,       
  top=1mm,         
  bottom=1mm       
]
\small
You are given several suspicious methods retrieved via embedding-based search. 
Your task is to carefully read each code snippet and determine how likely each method causes the bug described earlier. 
Then, **rank the methods** from most likely buggy (rank 1) to least likely buggy, output is in json form. \\

Use this JSON output schema:

method = \{'class': str, 'method':str, 'rank': int\}

return list[method]\\

class: \{class name\} 

method: \{method name\}

code snippet: \{method code\}    
\end{tcolorbox}




%% file: sections/experiment.tex
\section{Experimental Settings}

\phead{Benchmark.} We conduct our experiments on the Defects4J~\cite{defects4j} dataset, a widely used benchmark for evaluating fault localization techniques, which includes two stable versions: Defects4J-v1.2.0 and Defects4J-v2.0.0. The dataset consists of real-world Java projects with known bugs and corresponding test cases. 

In particular, we use Defects4J-v1.2.0 to compare \tool against baseline approaches. This version contains 395 bugs. During our evaluation, we exclude 14 bugs where the fault does not reside within any method body, as these represent edge cases outside the scope of method-level fault localization.
In addition, following prior studies~\cite{lou2021boosting,qinSoapFLStandardOperating2025}, we further evaluate the generalizability of \tool's performance using 226 additional bugs from Defects4J-v2.0.0 in Section~\ref{D4J2}.

\phead{Evaluation Metrics.} To measure the effectiveness of \tool, we employ the following metrics:

\uhead{Top-N} denotes the number of bugs for which the actual faulty method appears within the top N positions in the list.

\uhead{MAP (Mean Average Precision)} captures the average precision across all bugs, providing a comprehensive measure of how well \tool ranks all relevant methods in the list.

\uhead{MRR (Mean Reciprocal Rank)} quantifies the average rank of the first relevant result, showing how well \tool ranks the most relevant method in the list.


\phead{Configurations.} In RQ1, we adopt OpenAI's gpt-4.1-mini as the LLM component to ensure fair comparison with other baselines, together with Microsoft's UniXcoder as the code embedding model for semantic representation. For the subsequent analyses in RQ2, RQ3, and RQ4 we employ Google's gemini-2.0-flash as a more cost-efficient alternative, while still preserving the ability to assess the effectiveness of our framework. During retrieval, our framework obtains the top 40 results from Semantic Dense Retrieval and re-ranks them with the LLM to produce the final top 10 suspicious methods. The choice of 40 is empirically determined and further discussed in the sensitivity analysis of section~\ref{threats}.

%% file: sections/results.tex
\section{Results}

\noindent This section presents the results of our experiments by proposing and answering three research questions (RQs). 

\subsection{RQ1: How does \tool compare to state-of-the-art approaches?}

\phead{Motivation.} The recent advancements of LLMs in fault localization have attracted significant attention from the software engineering community. Therefore, in this RQ, we compare \tool with recent state-of-the-art approaches, both LLM-based and non-LLM, to assess its performance in accurately identifying faulty locations.

\phead{Approach.} We evaluate the effectiveness of \tool against state-of-the-art fault localization techniques from both LLM-based and non-LLM approaches. 
A survey study~\cite{kochharPractitionersExpectationsAutomated2016} with practitioners shows that most developers only inspect Top-5 elements during fault localization. Therefore, following prior work on fault localization~\cite{qinSoapFLStandardOperating2025,kangQuantitativeQualitativeEvaluation2024,lou2021boosting,sohn2017fluccs}, we evaluate the effectiveness in terms of Top-1, Top-3, and Top-5 accuracy. To ensure consistency, we employ gpt-4.1-mini as the base LLM for FaR-Loc and all LLM-based baselines. The selected baselines are as follows:




\uhead{LLM-based}: SoapFL~\cite{qinSoapFLStandardOperating2025}, AutoFL~\cite{kangQuantitativeQualitativeEvaluation2024}. SoapFL adopts a multi-agent framework that performs fault localization in two stages: file-level followed by method-level. AutoFL utilizes the function-calling capabilities of LLMs to autonomously navigate the codebase and identify buggy code. 

\uhead{Learning-based}: GRACE~\cite{lou2021boosting}, FLUCCS~\cite{sohn2017fluccs}. GRACE uses Gated Graph Neural Networks to leverage detailed coverage data and rank program entities. FLUCCS leverages learning-to-rank techniques to combine SBFL scores with code and change metrics. 

\uhead{Spectrum-based}: Ochiai~\cite{ochiai}. A classical SBFL approach that ranks suspicious elements using the Ochiai formula based on test failures and coverage data.

In addition, we conduct the cost analysis of \tool both in terms of API cost and time. For API cost, we calculate the token usage with the per-million-token price, specifically \$0.15 for input tokens and \$0.60 for output tokens~\cite{openai2024gpt4omini}.

\input{tables/RQ1_comparison}

\phead{Results.}~\textbf{\textit{\tool outperforms the other two LLM-based baselines SoapFL and AutoFL, by 14.6\% and 9.1\% in Top-1, by 19.2\% and 22.1\% in Top-5, respectively.}}
Table~\ref{tab:all-baselines-comparison} compares the results between \tool and the baseline techniques. Overall, \tool consistently outperforms all baselines with respect to all Top-N metrics. Compared to LLM-based techniques, \tool correctly locates 228 faults at Top-1, which represents a 14.6\% and 9.1\% improvement over SoapFL and AutoFL, respectively. \tool also demonstrates better effectiveness under the Top-3 and Top-5 metrics, achieving performance improvements ranging from 13.6\% to 22.1\%.
These results highlight the effectiveness of \tool's retrieval-augmented architecture, which consistently outperforms both SoapFL's multi-agent approach and AutoFL's function-calling strategy.



To better understand these results, we further investigate the performance of \tool on specific systems. We observe that its performance improves most significantly on Closure, which is also the most complex system in the benchmark.
\tool achieves 24.4\%, 19.0\%, and 26.9\% improvements compared to SoapFL, and 37.8\%, 82.9\%, and 107.3\% compared to AutoFL at Top-1, Top-3, and Top-5, respectively. 
We attribute the effectiveness of \tool on Closure to its integrated framework that combines focused functionality query generation with code embedding-based semantic retrieval. Prior studies~\cite{qinSoapFLStandardOperating2025,kangQuantitativeQualitativeEvaluation2024} found that localizing faults in complex systems like Closure is particularly challenging due to the large search space. For example, the average number of methods covered by failing tests in Closure is 634, compared to only 16 to 347 methods in other systems. 
Since the code coverage contains excessive information, it becomes difficult to identify the relevant methods among the many covered methods. To address this issue, \tool not only employs the LLM Functionality Extraction component to generate a concise query that captures the problematic functionality, but also applies semantic retrieval over pre-trained code embeddings to effectively reduce the reasoning space for the LLM.
The natural language description of the fault helps \tool focus on the most important information and avoids the risk of overwhelming it with unnecessary context.

\noindent\textbf{\textit{\tool also outperforms all learning-based and spectrum-based methods across every Top-N metric.}} To further evaluate the effectiveness of \tool, we compare it against non-LLM fault localization methods, including learning-based (GRACE and FLUCCS) and spectrum-based (Ochiai) techniques.
As shown in Table~\ref{tab:all-baselines-comparison},
\tool consistently outperforms all learning-based and spectrum-based techniques across all Top-N metrics and all studied systems.
\tool identifies 228 faults at Top-1, exceeding GRACE, FLUCCS, and Ochiai by 36, 68, and 148 faults, respectively.
Similarly, in terms of Top-3 and Top-5, \tool achieves an improvement between 3.4\% and 73.3\% over the other three baselines. 

The results demonstrate that LLM-based techniques can outperform learning-based techniques in zero-shot settings.
One key advantage of \tool is its ability to leverage pre-trained knowledge to accurately identify faults without requiring task-specific labeled data, using retrieval augmentation. Since LLMs are trained on massive corpora of code and natural language, they possess a deep understanding of language, allowing them to generate contextually relevant and semantically coherent links between suspicous code and failing functionality. In contrast, learning-based techniques rely on large labeled datasets, but their performance often drops on unseen systems. This advantage is particularly notable because learning-based techniques can leverage rich but potentially overfitted in-project knowledge. 
Our findings suggest that retrieval-augmented frameworks can effectively enhance general-purpose LLMs by focusing on relevant contextual information, establishing a new research direction for LLM-based fault localization approaches.


\noindent\textbf{\textit{\tool demonstrates competitive cost-effectiveness and efficiency.}}
Specifically, \tool costs only \$0.019 per bug compared to SoapFL's \$0.055 and AutoFL's \$0.065, which represents a cost reduction of approximately 65 to 70\%.
This significantly lower cost is expected as \tool only focuses on functionality-related context, which minimizes API calls and number of tokens being processed. 
Beyond monetary savings, \tool also achieves high computational efficiency. \tool uses open-source models on a single NVIDIA 2080Ti GPU with merely 728 MB of memory and computes embeddings for 100 methods in just 0.986 seconds. This efficiency leads to faster overall execution, and \tool can successfully localize 95\% of bugs within 60 seconds.

\rqboxc{RQ1 Takeaway}{\tool outperforms LLM-based baselines SoapFL and AutoFL, by 14.6\% and 9.1\% in Top-1, and by 19.2\% and 22.1\% in Top-5, respectively, while remaining comparably cost-effective. It also outperforms all learning-based and spectrum-based baselines across every Top-N metric.}

\subsection{RQ2: How do our design choices impact localization effectiveness?}\label{ablation}

\input{tables/RQ2_ablation_study}

\phead{Motivation.} In this RQ, we conduct an ablation study to investigate the contribution of each component individually and analyze its impact on the effectiveness of fault localization. The findings may inspire and provide insights for future work on partially adapting our design.

\phead{Approach.} We investigate the impact of key design choices on the fault localization performance of \tool. For each component, we conduct controlled experiments by removing or altering it while keeping the others unchanged, so as to isolate its contribution. This enables us to quantify the importance of each component and examine whether our framework contains unnecessary complexity or redundancy. To this end, we construct the following variants:

\uhead{w/o functionality query}
To evaluate the effectiveness of the functionality query, we use the raw inputs (i.e., test code and stack trace) directly as the retrieval query. 
In this setting, we further examine the effectiveness of incorporating the stack trace as textual context in query generation, and report that the quality of the functionality query has a substantial impact on fault localization performance.

\uhead{w/o semantic dense retrieval}
To evaluate the role of semantic dense retrieval, we remove the code embedding models and instead employ the traditional approach based on keyword matching BM25 algorithm~\cite{robertson2009probabilistic} to retrieve potentially faulty methods.

\uhead{w/o LLM re-ranking}
To assess the effectiveness of LLM re-ranking, we directly rely on the initial retrieval results from semantic dense retrieval and skip the re-ranking step.





\phead{Results.}~\textbf{\textit{LLM functionality extraction improves Top-5 accuracy by 11.2\%. Stack traces also provide valuable hints for LLM-based fault localization.}}
Using raw inputs (i.e., test code and stack trace) directly as the retrieval query leads to a noticeable performance degradation. Top-1 accuracy drops from 216 to 200, Top-3 from 286 to 258, and Top-5 from 307 to 276. Correspondingly, MAP decreases from 0.619 to 0.552, and MRR drops from 0.665 to 0.607. These results suggest that raw failure data may introduce noise that negatively impacts localization accuracy. Incorporating the failing functionality as a semantic query helps the retriever better filter relevant methods. Future studies should consider exploring beyond raw failure data when designing fault localization techniques.

In addition, removing the stack trace information also leads to a noticeable performance degradation: Top-1 accuracy drops from 216 to 195, Top-3 from 286 to 252, and Top-5 from 307 to 282. 
These results show that the presence of stack trace can significantly enhance the effectiveness of \tool by providing additional fault-relevant context (e.g., code elements related to the failing execution path). Our finding highlights the usefulness of stack trace in fault localization, which is consistent with prior studies~\cite{chen2019empirical,pacheco2024leveraging}.

\noindent\textbf{\textit{Semantic dense retrieval improves Top-5 accuracy by 18.5\%.}}
Replacing semantic dense retrieval with traditional keyword-based retrieval (BM25) leads to a significant performance drop. Specifically, Top-1 accuracy decreases from 216 to 186, Top-3 from 286 to 240, and Top-5 from 307 to 259. Correspondingly, MAP falls from 0.619 to 0.523, and MRR from 0.665 to 0.566. These results indicate that while keyword-based retrieval can help identify some fault-relevant methods, it is limited by its reliance on exact term matching and cannot capture deeper semantic relationships between the query and code components. In contrast, semantic dense retrieval uses pre-trained code representations to match codes and the intended functionality, rather than relying only on keyword overlap, resulting in a more reliable starting candidate list for re-ranking.

\noindent\textbf{\textit{LLM re-ranking significantly improves Top-5 accuracy by 24.8\%, playing a critical role in improving localization effectively by filtering out noisy candidates through semantic reasoning.}}
When LLM re-ranking is disabled, the Top-1 accuracy drops from 216 to 135, Top-3 from 286 to 214, and Top-5 from 307 to 246. Correspondingly, MAP decreases from 0.619 to 0.465, and MRR drops from 0.665 to 0.488. These results demonstrate that while the semantic dense retrieval component can help identify fault-relevant methods, the suspicious list (i.e., generated based on the similarity between the embeddings of covered methods and the failing functionality query) does not always reflect the root cause of the faults. The use of LLM-based re-ranking significantly improves the Top-5 accuracy by further filtering noisy candidates through deeper semantic reasoning.




\rqboxc{RQ2 Takeaway}{While all design choices contribute to the overall effectiveness of \tool, LLM-based re-ranking plays the most critical role by filtering noisy candidates through semantic reasoning, which significantly improves the Top-5 accuracy by 24.8\%.}

\subsection{RQ3: What is the effectiveness of \tool with different LLMs?}

\input{tables/RQ3_llm_comparison}


\phead{Motivation.}
Recent research~\cite{kangQuantitativeQualitativeEvaluation2024} has shown that the choice of LLM can significantly influence fault localization performance. In this RQ, we systematically evaluate the effectiveness and generalizability of \tool when integrated with different LLMs. Specifically, we examine how different LLMs affect functionality query generation quality and re-ranking accuracy. This analysis reveals how LLM selection impacts each component of our framework.

\phead{Approach.}
To evaluate the impact of different LLMs on \tool's performance, we integrate three representative models: gpt-4.1-mini, gemini-2.0-flash, and gpt-4o-mini. These models come from different providers and vary in capability and cost. We use this diversity to examine how LLM choice affects functionality query generation and re-ranking accuracy. We conduct our experiments on the Defects4J v1.2.0 benchmark, with Top-N (N=1, 3, 5), MAP, and MRR metrics. We report both intermediate retrieval and final re-ranking results to analyze the contribution of each LLM to the overall framework.

\phead{Results.}~\textbf{\textit{Retrieval results are consistently stable across LLMs, with differences ranging only from 1.4\% to 3.7\%.}}
Our experiments show that the choice of LLM for functionality query generation has trivial impact on the retrieval performance. 
As shown in Table~\ref{tab:llm-retrieval-reranking}, the number of relevant methods retrieved at Top-60 ranges from 347 (gpt-4.1-mini) to 352 (gemini-2.0-flash), with a mere 1.4\% difference. Similar consistency appears in the Top-20 and Top-40 results, where differences remain under 4.0\% (3.7\% and 3.6\%, respectively). Despite using different LLMs, functionality queries generated from the same failure information remain highly consistent.

We also invested how well the generated functionality queries actually match their corresponding buggy methods by measuring semantic similarity with UniXcoder.
The results show that the cosine similarity scores are consistent across all LLMs: 0.866 for gpt-4.1-mini, 0.890 for gemini-2.0-flash, and 0.860 for gpt-4o-mini. 
These results indicate that all tested LLMs generate functionality queries that align well with the target code, and thereby maintain high retrieval performance.

\noindent\textbf{\textit{The reasoning ability of the LLM is crucial for effective re-ranking.}}
In contrast, the choice of LLM has a much greater impact on re-ranking performance. As shown in Table~\ref{tab:llm-retrieval-reranking}, gpt-4.1-mini achieves the highest Top-1, Top-3, and Top-5 accuracies (228, 284, and 304, respectively), as well as the best MAP (0.623) and MRR (0.679). By comparison, gpt-4o-mini performs noticeably worse, with Top-1 accuracy dropping to 180 (21.1\% lower) and MAP to 0.520 (16.5\% lower). Overall, stronger LLMs improve re-ranking effectiveness by up to 26.7\% in Top-1 accuracy and 19.8\% in MAP compared to weaker models.
These results suggest that while functionality query generation ensures strong retrieval, the reasoning capabilities of the LLM are crucial for effective re-ranking and higher fault localization accuracy. We also observe that although gemini-2.0-flash performs best in retrieval results, its re-ranking performance is relatively weaker than gpt-4.1-mini in the Top-1 metric.
Our findings suggest that future studies should consider using different models for different parts of the pipeline.


\rqboxc{RQ3 Takeaway}{\tool's retrieval performance remains stable across LLMs, but its re-ranking accuracy varies significantly. Stronger LLMs improve Top-1 accuracy by up to 26.7\%, underscoring the importance of reasoning ability in FL.}

\subsection{RQ4: What is the effectiveeness of \tool with different code embedding models?}


\phead{Motivation.} A prior study~\cite{ding2022can} finds that different code embeddings can have different impact on the performance of downstream software engineering tasks. Therefore, in this RQ, we investigate the generalizability of \tool across different embedding models.

\phead{Approach.} The choice of embedding model plays an important role in determining the quality of semantic matching between failing functionality descriptions and potentially faulty code components. As such, it directly affects the initial retrieval precision and, consequently, the overall localization performance. Therefore, we integrate three representative models into our framework: CodeBERT, CodeT5+, and UniXcoder. Each model employs different architectures and pre-training approaches for code representation. 

\input{tables/RQ4_embedding_models_comparison}

\phead{Results.}~\textbf{\textit{UniXcoder significantly outperforms other embedding models for fault localization across all Top-N accuracy.}} Table~\ref{tab:embedding-model} compares the fault localization performance of \tool when using different code embedding models for retrieval. Retrieval performance varies greatly across embedding models, which directly determines the upper bound of the final re-ranking results. Among the evaluated models, UniXcoder yields the best results across all evaluation metrics. Specifically, UniXcoder locates 216 faults at Top-1 accuracy, substantially outperforming CodeT5+ and CodeBERT by 11.3\% and 49.0\%, respectively. The MAP and MRR scores further confirm this trend, with UniXcoder achieving 0.619 and 0.665 respectively, representing relative improvements of 15.0\% (MAP) and 15.2\% (MRR) over CodeT5+, and 63.7\% (MAP) and 58.3\% (MRR) over CodeBERT.
These results suggest that UniXcoder provides more accurate and semantically relevant code representations, making it better suited for fault localization tasks.

\begin{figure}[tb]
  \centering
  \includegraphics{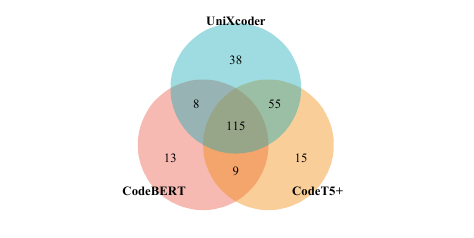}
  \caption{\tool's overlap results between different embeddings.}
  \label{fig:overlap}
\end{figure}

To further understand the behavior of different embeddings, we analyze the overlap of Top-1 results across UniXcoder, CodeT5+, and CodeBERT. As shown in Figure~\ref{fig:overlap}, although there is a intersection among the methods correctly identified by each embedding, each model also contributes uniquely localized bugs. This observation highlights the complementary nature of embedding models—while UniXcoder is overall superior, CodeT5+ and CodeBERT can still successfully localize some bugs that UniXcoder misses. This motivates future work to consider ensemble or hybrid retrieval strategies that leverage the strengths of multiple encoder architectures.

\rqboxc{RQ4 Takeaway}{The choice of embedding model greatly affects \tool's performance. Specifically, UniXcoder exhibits the best performance. When integrating or applying \tool, practitioners and future studies may consider ensemble or hybrid retrieval strategies to benefit from multiple encoder architectures.}

%% file: tables/RQ1_comparison.tex
\begin{table*}[tb]
 \caption{Comparison between \tool, LLM-based, learning-based, and spectrum-based baselines on Defects4J-v1.2.0 (gpt-4.1-mini)}
 \begin{center}
 \resizebox{\textwidth}{!}{
 \begin{tabular}{l|c|ccc|ccc|ccc|ccc|ccc|ccc}
 \toprule
 \multirow{3}{*}{\textbf{Systems}} & \multirow{3}{*}{\textbf{\# Bugs}} 
 & \multicolumn{9}{c|}{\textbf{LLM-based (gpt-4.1-mini)} }
 & \multicolumn{6}{c|}{\textbf{Learning-based} }
 & \multicolumn{3}{c}{\textbf{{Spectrum-based}}} \\
 \cline{3-20}
 & & \multicolumn{3}{c|}{FaR-Loc} 
   & \multicolumn{3}{c|}{SoapFL} 
    & \multicolumn{3}{c|}{AutoFL}
    & \multicolumn{3}{c|}{GRACE} 
    & \multicolumn{3}{c|}{FLUCCS} 
    & \multicolumn{3}{c}{Ochiai} \\
  [-0.2ex]
  & & Top1 & Top3 & Top5 & Top1 & Top3 & Top5 & Top1 & Top3 & Top5 
    & Top1 & Top3 & Top5 & Top1 & Top3 & Top5 & Top1 & Top3 & Top5 \\
  \midrule
  Chart   & 25  & 17 & 21 & 22 & 18 & 23 & 23 & 21 & 23 & 23 & 14 & 20 & 22 & 15 & 19 & 16 & 6  & 14 & 15 \\
  Closure & 130 & 51 & 75 & 85 & 41 & 63 & 67 & 37 & 41 & 41 & 47 & 70 & 81 & 42 & 66 & 77 & 14 & 30 & 38 \\
  Lang    & 61  & 54 & 58 & 60 & 45 & 51 & 51 & 52 & 58 & 58 & 42 & 54 & 57 & 40 & 53 & 55 & 24 & 44 & 50 \\
  Math    & 104 & 72 & 87 & 89 & 62 & 74 & 75 & 67 & 86 & 87 & 61 & 78 & 89 & 48 & 77 & 83 & 23 & 52 & 62 \\
  Mockito & 36  & 19 & 24 & 28 & 20 & 23 & 23 & 17 & 22 & 22 & 17 & 24 & 26 & 7  & 19 & 22 & 7  & 14 & 18 \\
  Time    & 25  & 15 & 19 & 20 & 13 & 16 & 16 & 15 & 18 & 18 & 11 & 14 & 19 & 8  & 15 & 18 & 6  & 11 & 13 \\
  \midrule
  Overall & 381 & \textbf{228} & \textbf{284} & \textbf{304} 
          & 199 & 250 & 255 & 209 & 248 & 249
          & 192 & 260 & 294 & 160 & 249 & 271 & 80  & 165 & 196 \\
  \bottomrule
  \end{tabular}
  }
  \label{tab:all-baselines-comparison}
  \end{center}
\end{table*}

%% file: tables/RQ2_ablation_study.tex
\begin{table*}[tb]
  \caption{Ablation study of \tool on Defects4J-v1.2.0 (gemini-2.0-flash)}
  \begin{center}
  \begin{tabular}{l|ccc|cc}
  \toprule
  \textbf{Variants} & \textbf{Top1} & \textbf{Top3} & \textbf{Top5} & \textbf{MAP} & \textbf{MRR} \\
  \midrule
  \tool        & \textbf{216} & \textbf{286} & \textbf{307} & \textbf{0.619} & \textbf{0.665} \\
  w/o functionality query  & 200 & 258 & 276 & 0.552 & 0.607 \\
  w/o code embeddings (BM25)      & 186 & 240 & 259 & 0.523 & 0.566 \\
  w/o LLM re-ranking       & 135 & 214 & 246 & 0.465 &	0.488 \\
  \bottomrule
  \end{tabular}
  \label{tab:ablation-study}
  \end{center}
\end{table*}

%% file: tables/RQ3_llm_comparison.tex
\begin{table*}[tb]
 \caption{FL performance with different LLMs on Defects4J-v1.2.0}
 \begin{center}
 \begin{tabular}{l|ccc|ccccc}
 \toprule
 \multirow{2}{*}{\textbf{LLM}} 
 & \multicolumn{3}{c|}{\textbf{Retrieval Results} }
 & \multicolumn{5}{c}{\textbf{Re-ranking Results}} \\
 & Top-20 & Top-40 & Top-60 & Top-1 & Top-3 & Top-5 & MAP & MRR \\
 \midrule
 gpt-4.1-mini   & 300 & 332 & 347 & \textbf{228} & \textbf{284} & \textbf{304} & \textbf{0.623} & \textbf{0.679} \\
 gemini-2.0-flash & \textbf{309} & \textbf{341} & \textbf{352} & 216 & 286 & 307 & 0.619 & 0.665 \\
 gpt-4o-mini    & 298 & 329 & 348 & 180 & 243 & 261 & 0.520 & 0.566 \\
 \bottomrule
 \end{tabular}
 \label{tab:llm-retrieval-reranking}
 \end{center}
\end{table*}

%% file: tables/RQ4_embedding_models_comparison.tex
\begin{table*}[tb]
   \caption{FL performance with different embedding models on Defects4J-v1.2.0 (gemini-2.0-flash)}
  \begin{center}
  \begin{tabular}{l|ccc|ccccc}
  \toprule
  \multirow{2}{*}{\textbf{Embedding Model}} 
   & \multicolumn{3}{c|}{\textbf{Retrieval Results}} 
   & \multicolumn{5}{c}{\textbf{Re-ranking Results}} \\
   & Top-20 & Top-40 & Top-60 & Top-1 & Top-3 & Top-5 & MAP & MRR \\
  \midrule
  UniXcoder  & \textbf{309} & \textbf{341} & \textbf{352} & \textbf{216} & \textbf{286} & \textbf{307} & \textbf{0.619} & \textbf{0.665} \\
  CodeT5+    & 218 & 271 & 290 & 194 & 237 & 258 & 0.539 & 0.577 \\
  CodeBERT   & 147 & 191 & 213 & 145 & 173 & 181 & 0.379 & 0.420 \\
  \bottomrule
  \end{tabular}
  \label{tab:embedding-model}
  \end{center}
\end{table*}

%% file: sections/discussion.tex
\section{Discussion}




\subsection{Generalizability of \tool on Additional Datasets}\label{D4J2}

\input{tables/d4j_v2.tex}

\noindent To evaluate generalizability, we follow prior studies~\cite{lou2021boosting,xuFlexFLFlexibleEffective2024} and assess \tool on 226 additional faults from the Defects4J-v2.0.0 benchmark. Table~\ref{tab:generalizability evaluation} presents the effectiveness of \tool, SoapFL, GRACE and Ochiai. \tool outperforms all baseline methods across the Top-N metrics. 
In particular, \tool localizes 122 bugs within Top-1, outperforming SoapFL (99), GRACE (85), and Ochiai (32). The Top-1 improvement over GRACE (i.e., 43.5\%) is even higher than that observed on Defects4J-V1.2.0 (i.e., 18.8\%).
Moreover, the improvements persist in the Top-3 and Top-5 metrics, where \tool locates 159 and 167 faults, respectively. These results suggest that \tool performs even better on newer and more diverse studied systems, highlighting its generalizability across different benchmarks.

\subsection{Case Study and Implications}

\noindent To demonstrate the practical implications of our approach, we conduct a case study to showcase the effectiveness of \tool in addressing two major challenges in fault localization. First, LLM-based fault localization often struggles to locate faults in large search space (i.e., the entire code software system). Although prior studies~\cite{qinSoapFLStandardOperating2025,xuFlexFLFlexibleEffective2024,wuLargeLanguageModels2023} have proposed methods for automatically identifying fault-relevant methods using LLMs, their overall effectiveness often falls short compared to learning-based approaches. For example, SoapFL's Top-1 metric on Closure, the largest project in Defects4J, drops to 31.5\%, significantly lower than its average performance on other systems, 62.9\%.
Second, while many LLM-based methods~\cite{qinFaultLocalizationSemantic2024,xuFlexFLFlexibleEffective2024,kangQuantitativeQualitativeEvaluation2024, rafi2024enhancing} attempt to leverage the failing test information (e.g., test code and stack trace), they directly use this information as context to improve localization performance. Yet, such information often contains limited or irrelevant information for fault diagnosis (e.g., third-party method calls or code executed after the failure point). Therefore, to investigate the effectiveness of \tool in addressing theses challenges, we showcase a fault (i.e., \textit{Closure-112}) that contains limited diagnostic information from a large and complex system. 


\begin{figure*}[t] 
  \centering
  \includegraphics[width=\textwidth]{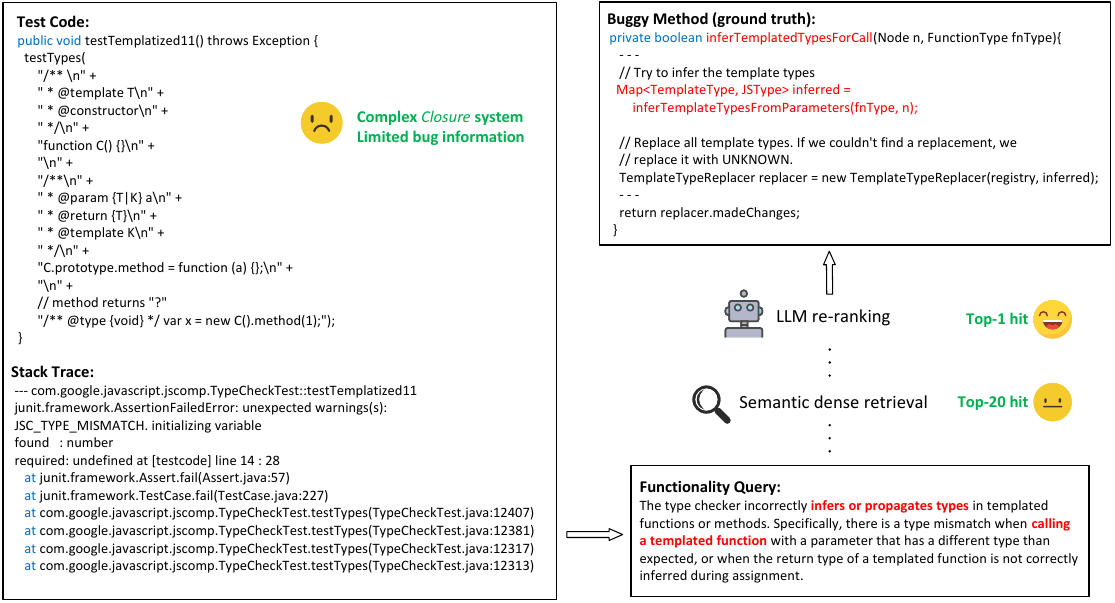}
  \caption{Case Study of \tool on Closure-112.}
  \label{fig:case_study}
\end{figure*}

Figure~\ref{fig:case_study} presents the details of \textit{Closure-112}, including the test code, stack trace, buggy method and the functionality query generated by \tool. By combining the LLM’s natural language understanding capabilities with the code comprehension strength of embedding models, \tool successfully localizes the buggy method at Top-1, effectively addressing both identified challenges. In comparison, SoapFL locates the fault beyond Top-5.

\phead{Effectiveness of Functionality Query.}
As illustrated in Figure~\ref{fig:case_study}, both the test code and stack trace provide limited diagnostic information for localizing the buggy method. The test code only invokes a helper function, \texttt{testTypes}, while the stack trace presents only a partial execution path up to the assertion failure, omitting intermediate method calls. \tool addresseses this challenge by first reasoning about the failing functionality. It identifies that the issue lies in the \textit{type checker}, which \textit{incorrectly infers or propagates types in templated functions}. More specifically, it determines that  \textit{there is a type mismatch when calling a templated function}, as highlighted in the functionality query shown in Figure~\ref{fig:case_study}.
This insight helps guide \tool's semantic search for relevant methods.
Notably, keywords such as \textit{infers} and \textit{calling} do not appear in the test code or stack trace, yet are highly relevant to the faulty method \texttt{inferTeamplatedTypesForCall} due to their semantic similarity. 
The functionality query is a contributing factor to the effectiveness of \tool. Without it, test code and stack trace alone contain too much noise, which may hinder the accurate retrieval and localization of fault-relevant methods. In the absence of the functionality query, \tool places the faulty method outside the Top-5. With this semantic clue, however, \tool re-rank it to Top-1.

\uhead{Implication.} 
Incorporating failing functionality queries enhances FL by generating additional semantic context that goes beyond the information available in test code and stack traces. 
Our results show that \tool achieves significant improvements when reasoning about failing functionality. 
Future studies should explore the systematic integration of failing functionality insights to further improve the effectiveness of FL techniques.


\begin{table}[tb]
  \caption{Comparison of Code Embedding Models}
  \label{tab:model-comparison}
  \centering
  \renewcommand{\arraystretch}{1.6}
  \setlength{\tabcolsep}{5pt}
  \begin{tabular}{cccc}
    \toprule
    \textbf{Model} & \textbf{Year} & \textbf{Architecture} & \textbf{Pre-training Data} \\
    \midrule
    CodeBERT & 2020 & Encoder-only & \makecell{CodeSearchNet} \\
    \midrule
    UniXcoder & 2022 & Unified Transformer & \makecell{CodeSearchNet,\\ ASTs} \\
    \midrule
    CodeT5+ & 2023 & Encoder-decoder (T5) & \makecell{CodeSearchNet, \\ GitHub Code Dataset} \\
    \bottomrule
  \end{tabular}
\end{table}

\phead{Effectiveness of Embedding Models.}
We further evaluate the impact of different code embedding models (CodeBERT, UniXcoder, and CodeT5+) on Closure-112. While all three models are well-established for code representation tasks, their effectiveness varies considerably in our framework. 
In particular, our framework with UniXcoder successfully ranks the faulty method at Top-1, whereas those with CodeBERT and CodeT5+ place it outside the Top-5. This performance gap may be attributed to differences in their pre-training process.
Table~\ref{tab:model-comparison} summarizes the key characteristics of the three code embedding models, including their architectural design and pre-training corpora. 
Notably, our framework with UniXcoder outperforms those with CodeT5+ and CodeBERT across all localization metrics, likely due to its distinctive use of Abstract Syntax Trees (ASTs) during pre-training \cite{guoUniXcoderUnifiedCrossModal2022}.
The structural knowledge encoded by ASTs enhances UniXcoder's ability to capture both code semantics and syntax structure, which is particularly beneficial for fault localization tasks.

Interestingly, CodeT5+, despite being trained on a substantially larger dataset with token counts reportedly 50 times greater than CodeSearchNet \cite{wangCodeT5OpenCode2023}, performs worse than UniXcoder. This suggests that merely scaling up training data without incorporating structural signals may be insufficient for tasks that benefit from fine-grained code understanding.

\uhead{Implication.}
The effectiveness of different code embedding models in fault localization varies considerably. Models that incorporate structural code information—such as Abstract Syntax Trees (ASTs) in UniXcoder—consistently outperform those that rely solely on large-scale pre-training data. This suggests that future FL approaches should further exploit structural code representations to enhance embedding quality.

\phead{Effectiveness of LLM Re-ranking.}
As demonstrated in our earlier ablation study, LLM re-ranking plays a crucial role in our framework. This is further evidenced in the Closure-112 case study. The buggy method \texttt{inferTemplatedTypesForCall} does not always appear at the top of the initial candidate list produced by semantic dense retrieval. However, after applying LLM re-ranking the correct method is consistently ranked at Top-1. We attribute this to the fact that semantic dense retrieval may introduce noise, resulting in irrelevant methods being included in the candidate list. That been said, the LLM re-ranking module can compensate for this by analyzing the code and performing deeper reasoning, which allows \tool to accurately prioritize the actual buggy method.

\uhead{Implication.} 
LLM re-ranking effectively mitigates the noise introduced by dense retrieval, leveraging its code understanding and reasoning capabilities to improve the accuracy of fault localization.
This suggests that future studies should further explore specialized re-ranking strategies, or similar hybrid designs that combine lightweight retrieval techniques with reasoning-intensive LLMs. 


%% file: tables/d4j_v2.tex
\begin{table}[tb]
  \caption{Comparison of Fault Localization Techniques on Defects4J-v2.0.0 (gpt-4.1-mini)}
  \begin{center}
  \begin{tabular}{c|c|l|ccc}
    \toprule
  \textbf{Project} & \textbf{\# Bugs} & \textbf{Techniques} & \textbf{Top1} & \textbf{Top3} & \textbf{Top5} \\
    \midrule
    \multirow{4}{*}{Overall} 
      & \multirow{4}{*}{214} 
      & \textbf{FaR-Loc}         & \textbf{122} & \textbf{159} & \textbf{167} \\
      &                 & SoapFL       & 99           & 128           & 134           \\
      &                 & GRACE        &  85           & 119           & 140           \\
      &                 & Ochiai       &  32           &  74           &  93           \\
    \bottomrule
  \end{tabular}
  \label{tab:generalizability evaluation}
  \end{center}
\end{table}

%% file: sections/threats.tex
\section{Threats to Validity}\label{threats}

\phead{Internal Validity.}
One potential threat to internal validity is data leakage. Since the Defects4J dataset may be included in the training data of LLMs, there is a risk that the models could inadvertently access information about the defects under study. To mitigate this risk, we ensure that our framework does not expose any project names, identifiers, or human-provided labels during the evaluation process.

\phead{External Validity.}
A threat to external validity concerns the generalizability of our results. Our experiments are primarily conducted on the Defects4J-v1.2.0 dataset, which may not fully represent all real-world software projects. To address this threat, following prior studies~\cite{qinSoapFLStandardOperating2025,lou2021boosting}, we further evaluate \tool on 226 additional faults from Defects4J-v2.0.0, and the consistent results across versions support the generalizability of our findings. Additionally, our current evaluation is limited to Java projects. Future work is needed to assess the applicability of our approach to other programming languages.

\phead{Sensitivity Analysis.}
We conduct a sensitivity analysis to examine the impact of varying the number of retrieved results provided to the LLM re-ranking component. Specifically, we experiment with supplying the top 20, 40, and 60 retrieval results. The performance remains similar, with only a 3.36\% variation across these settings. We use 40 retrieved results in our approach, as it yields the best outcome.


%% file: sections/conclusion.tex
\section{Conclusion}
\noindent In this paper, we propose \tool, a novel fault localization framework that integrates LLMs with RAG.
\tool introduces three components: LLM Functionality Extraction, Semantic Dense Retrieval, and LLM Re-Ranking, which collectively advance the effectiveness of fault localization.
On the Defects4J-v1.2.0 benchmark, \tool significantly outperforms state-of-the-art LLM-based, learning-based and spectrum-based baselines, achieving consistent Top-1 improvements. 
The API usage and runtime cost of \tool are comparable to those of other LLM-based methods, making it a practical choice for fault localization.
Our ablation study shows that each component contributes to the overall performance of \tool.
Further analysis reveals that the choice of models plays a important role in its effectiveness.
While functionality queries generated by different LLMs are largely consistent, their effectiveness in re-ranking suspicious methods varies considerably. Similarly, code embedding models that incorporate Abstract Syntax Trees (ASTs), such as UniXcoder, achieve greater improvements and highlight the importance of structural representation in fault localization.
To the best of our knowledge, \tool is the first fault localization approach that leverages RAG framework with strong performance.
Our findings suggest that integrating LLMs with RAG and functionality-aware design can bridge the gap between general-purpose language models and the challenges of real-world fault localization.